\documentclass[aps,twocolumn,showpacs,floatfix,prl]{revtex4}

\usepackage{graphicx}
\usepackage{dcolumn}
\usepackage{bm}
\usepackage{epsf}
\usepackage{subfigure}
\usepackage{epstopdf}
\usepackage{amsmath}
\usepackage{amssymb}

\newcommand{\e}{{\rm e}}

\begin{document}

\title{Universality of efficiency at maximum power}

\author{Massimiliano Esposito}
\altaffiliation[]{Also at Center for Nonlinear Phenomena and Complex Systems,
Universit\'e Libre de Bruxelles, Code Postal 231, Campus Plaine, B-1050 Brussels, Belgium.\\}
\author{Katja Lindenberg}
\affiliation{Department of Chemistry and Biochemistry and Institute for Nonlinear Science, 
University of California, San Diego, La Jolla, CA 92093-0340, USA}
\author{Christian Van den Broeck}
\affiliation{Hasselt University, B-3590 Diepenbeek, Belgium}

\date{\today}

\begin{abstract}
We investigate the efficiency of power generation by thermo-chemical engines. For strong coupling between the particle and heat flows and in the presence of a left-right symmetry in the system, we demonstrate that the efficiency at maximum power displays universality up to quadratic order in the deviation from equilibrium. A maser model is presented to illustrate our argument.
\end{abstract}



\maketitle


The concept of Carnot efficiency is a cornerstone of thermodynamics. It states that the efficiency of a
cyclic (``Carnot") thermal engine that transforms an amount $Q_r$ of energy extracted from a heat reservoir
at temperature  $T_r$ into an amount of work $W$ is at most $\eta=W/Q_r\leq\eta_c=1- T_l/T_r$, where
$T_l$  is the temperature of a second, colder reservoir. The theoretical implications of this result are
momentous, as they lie at the basis of the introduction by Clausius of the entropy as a state function.
The practical implications are more limited, since the upper limit $\eta_c$ (``Carnot efficiency") is
only reached for engines that operate reversibly. As a result, when the efficiency is maximal, the output
power is zero.  By optimizing the Carnot cycle with respect to power rather than efficiency,  Curzon
and Ahlborn found that the corresponding efficiency is given by $\eta_{CA}=1-\sqrt{ T_l/T_r}$~\cite{curzon}. 
They obtained this result for a specific model, using in addition the so-called endo-reversible
approximation (i.e., neglecting the dissipation in the auxiliary, work producing entity). Subsequently,
the validity of this result as an upper bound, as well as its universal character, were the subject
of a longstanding debate.  In the regime of linear response, more precisely to linear order
in $\eta_{c}$, it was proven that the efficiency at maximum power is indeed limited by
the Curzon-Ahlborn efficiency, which in this regime is exactly half of the Carnot efficiency,
$\eta\leq\eta_{CA}=\eta_c/2+O(\eta_c^2)$~\cite{VandenBroeck05}. The upper limit is reached for a
specific class of models, namely, those for which the heat flux is strongly coupled to the work-generating
flux. Interestingly,  such strong coupling is also a prerequisite for open systems to achieve Carnot
efficiency~\cite{kedem,VandenBroeck6}. In the nonlinear regime, no general result is known.
Efficiencies at maximum power, not only below but also above Curzon-Ahlborn efficiency, have
been reported~\cite{allamajan,japanese,Seifert07b,EspositoLindWdB}. However,  it was also found, again
in several strong coupling models~\cite{Seifert07b,EspositoLindWdB,Tu}, that the efficiency at maximum
power agrees with $\eta_{CA}$ up to quadratic order in $\eta_c$, i.e., $\eta=\eta_c/2+\eta_c^2/8+O(\eta_c^3)$,
again raising the question of universality at least to this order. In this letter we prove that the
coefficient $1/8$ is indeed universal for strong coupling models that possess a left-right
symmetry.  Such a universality is remarkable in view of the fact that most explicit macroscopic
relationships, for example the symmetry of Onsager coefficients, are limited to the regime
of linear response. The interest in strong coupling is further motivated by the observation that it
can naturally be achieved in nano-devices~\cite{Prost99,Linke,Cleuren}.
To complement our theoretical discussion, we also present a detailed study  of a
thermal nano-machine based on the operation of a maser~\cite{Scovil59}. It can be solved analytically
and illustrates all the above mentioned features. Depending on the value of the Einstein coefficients,
the efficiency of the maser at maximum power may be above or below Curzon-Ahlborn. However, when the
Einstein coefficients \emph{are equal}, the predicted universality is observed,
with the universal value $1/2$ for the linear coefficient, and the quadratic coefficient equal to $1/8$. 

In view of the interest of our analysis for small scale systems, and in order to establish the
connection with the subsequent discussion of the maser model, we derive the main results
on the basis of a stochastic thermodynamic analysis as formulated for a master equation description
of a driven open system~\cite{Schnakenberg,Esposito07}. As we will show in passing, this formalism
is fully consistent with macroscopic thermodynamics. 

The system under consideration is characterized by a set of states $i$ of energy $\epsilon_{i}$ and number
of particles $N_{i}$.  It exchanges particles and energy with two reservoirs $\mu=l,r$, with
inverse temperatures $\beta_{\mu}$ and chemical potentials $\mu_{\mu}$, respectively. 
The probability of finding the system in state $i$ at time $t$ is denoted by $p_i(t)$. 
The state of the system evolves in time according to a stochastic process
which is described by the master equation
\begin{eqnarray}
\dot{p}_i(t) = \sum_{j} W_{ij} p_{j}(t). \label{MEq}
\end{eqnarray}
As a result of conservation of total probability, the stochastic rate matrix obeys
the usual condition $\sum_{i} W_{ij}=0$. $W_{ij}$ is the probability per unit time to make a transition
to state $i$ from state $j$. We assume that these transition rates are expressed as sums of
independent contributions from the two reservoirs, $W_{ij} = \sum_{\nu} W_{ij}^{(\nu)}$  ($\nu=l,r$).
To reproduce the correct properties at equilibrium, it follows that each of the separate rate matrices
$W_{ij}^{(\nu)}$ satisfies detailed balance with respect to the grand canonical distribution at
the prevailing temperature and chemical potential,
\begin{eqnarray}
\frac{W_{ji}^{(\nu)}}{W_{ij}^{(\nu)}} = \exp{ \bigg\{ \beta_{\nu} \big[ 
\big(\epsilon_i -\epsilon_{j}\big)-\mu_{\nu} (N_{i}-N_{j}) \big] \bigg\}}.
\label{MEqaaaaa}
\end{eqnarray}
The average energy and matter currents entering the system from the reservoir $\nu$ are given by
\begin{eqnarray} 
&&{\cal I}_{E}^{(\nu)}(t) = \sum_{i,j} W_{ij}^{(\nu)} p_{j}(t) (\epsilon_{i} -\epsilon_{j}), \label{MEqaaaae}\\
&&{\cal I}_{M}^{(\nu)}(t) = \sum_{i,j} W_{ij}^{(\nu)} p_{j}(t) (N_{i} -N_{j}).
 \label{MEqaaaaf}
\end{eqnarray}
The rate of change of the total energy of the system and the (chemical) work per unit time on
the system read
\begin{eqnarray} 
&&\dot{{\cal E}}(t)= \sum_{i} \dot{p}_i(t) \epsilon_i=\sum_{\nu}  {\cal
I}_{E}^{(\nu)}(t),\nonumber\\
&&\dot{{\cal W}}(t)= \sum_{\nu} \mu_{\nu} {\cal I}_{M}^{(\nu)}(t) .\label{MEqaaaafc}
\end{eqnarray}
The corresponding total average heat flow follows from energy conservation, $
\dot{{\cal Q}}(t)= \sum_{\nu} \dot{{\cal Q}}^{(\nu)}(t) = \dot{{\cal E}}(t)- \dot{{\cal W}}(t) $.
In particular, the (average) heat flow from the reservoir $\nu$ into the system is given by 
\begin{eqnarray} 
\dot{{\cal Q}}^{(\nu)}(t) = {\cal I}_{E}^{(\nu)}(t) - \mu_{\nu} {\cal I}_{M}^{(\nu)}(t). \label{MEqaaaad}
\end{eqnarray}
The entropy of the system is taken to be the usual system entropy $S(t)= - \sum_m p_m(t) \ln p_m(t)$ (Boltzmann's constant $k_B=1$). 
Using the master equation~(\ref{MEq}), one  easily verifies that the rate of change of this entropy can be 
written in the form of  a balance equation, namely, $\dot{S}(t) = \dot{S}_i(t) + \dot{S}_e(t)$.
Here, $\dot{S}_i(t)$ is the non-negative total entropy production for the physical processes represented
by the master equation,
\begin{eqnarray} 
 \dot{S}_i(t) = \sum_{i,j,\nu} W_{ij}^{(\nu)} p_{j}(t) 
\ln \frac{W_{ij}^{(\nu)} p_{j}(t)}{W_{ji}^{(\nu)} p_{i}(t)} \geq 0.
\label{MEqaaaab}
\end{eqnarray}
Using Eq.~(\ref{MEqaaaaa}), one verifies that the entropy flow into the system is given by the familiar 
thermodynamic expression in terms of the heat fluxes, $\dot{S}_e(t) =\sum_{\nu} \dot{{\cal Q}}^{(\nu)}(t)/T_{\nu}$.

We focus on the case of a nonequilibrium steady state. From $\dot{S}=0$
we have that $\dot{S}_i =-\dot{S}_e$.
Also, current conservation at steady-state implies that $\sum_{\nu} {\cal I}_{E}^{(\nu)}
= \sum_{\nu} {\cal I}_{M}^{(\nu)} = 0.$
As a result, the entropy production can now be written in the traditional bilinear force-flux form~\cite{GrootMazur}
\begin{eqnarray} 
\dot{S}_i  = {\cal F}_{E} {\cal I}_{E}^{(r)} + {\cal F}_{M} {\cal I}_{M}^{(r)} \geq 0,
\label{MEqaaaai}
\end{eqnarray}
with the standard expressions for the thermodynamic forces,
\begin{eqnarray} 
{\cal F}_{E} = \frac{1}{T_{l}} - \frac{1}{T_{r}} ,~~
{\cal F}_{M} = (-\frac{\mu_{l}}{T_{l}}) - (-\frac{\mu_{r}}{T_{r}}) .\label{MEqaaaak}
\end{eqnarray}

We are interested in  the operation of the device as a  heat engine that carries
particles uphill in the chemical potential, driven by a heat current from the hot 
to the cold reservoir. With no loss of generality, we  henceforth assume that $T_r>T_l$ and $\mu_r<\mu_l$.
As mentioned before, we focus on the power generated by the device, which, due to current conservation
in the steady-state, reads:
\begin{eqnarray} 
{\cal P}=-\dot{{\cal W}}= -(\mu_{r}-\mu_{l}) {\cal I}_{M}^{(r)} .
\label{MEqaaaal}
\end{eqnarray}
The resulting efficiency of producing chemical work from the heat pumped out of the hot reservoir $r$
is [see Eq.~(\ref{MEqaaaad})]
\begin{eqnarray}
\eta =\frac{-{\cal W}}{{\cal Q}^{(r)}}= \frac{-\dot{{\cal W}}}{\dot{{\cal Q}}^{(r)}}
= \frac{\mu_{r}-\mu_{l}}{\mu_{r}-{\cal I}_{E}^{(r)}/{\cal I}_{M}^{(r)}}.
\label{MEqaaaam}
\end{eqnarray}

The above formalism can be further simplified for the case of \emph{strong coupling} between the energy
and matter flux, defined as 
\begin{equation}
{\cal I}_{E}^{(r)} = \varepsilon {\cal I}_{M}^{(r)} \equiv \varepsilon {\cal I}. \label{StrCouplGen}
\end{equation}
This condition implies that  energy is  exclusively transported by particles of a given
energy $\varepsilon$. Such a selection is quite natural in quantum nano-devices such as the maser~\cite{Scovil59}
(see also below), and in thermo-electrical nano-devices~\cite{Linke}, but it can also occur in the classical
context, for example for Kramers' escape, where the particles that cross the barrier have
precisely the minimum energy needed to do so~\cite{Tu}. 
Using Eq.~(\ref{StrCouplGen}), one can now rewrite the entropy production~(\ref{MEqaaaai}) in the simple form
$ \dot{S}_i  = {\cal F} {\cal I} \label{EntropyProdSCGen}$.
Here ${\cal I}$ is the current introduced in Eq.~(\ref{StrCouplGen}) and the associated thermodynamic force
${\cal F}$ can be expressed in terms of dimensionless scaled energies $x_l$ and $x_r $,
\begin{eqnarray}
{\cal F} = x_l-x_r  ; \  \ \  \ x_r = \frac{\varepsilon-\mu_r}{T_r} ,\ \ \  \ 
x_l = \frac{\varepsilon-\mu_l}{T_l}. \label{DefOfXSCGen}
\end{eqnarray}
Hence the two flows and forces collapse into a single flux  ${\cal I}$ and a single corresponding
thermodynamic force ${\cal F} $, respectively. Note that equilibrium, that is, zero entropy production,
is reached for ${\cal F}=0$. This does not require that the forces ${\cal F}_E$ and ${\cal F}_M$ be zero
separately. In fact in the vicinity of ${\cal F}=0$ the device can operate at Carnot efficiency, see
for example~\cite{kedem,VandenBroeck6,Linke}.
Using Eq.~(\ref{StrCouplGen}), the power (\ref{MEqaaaal}) becomes
\begin{eqnarray}
{\cal P}=-\dot{{\cal W}} = ( T_r x_r- T_l x_l) {\cal I}=-T_r ({\cal F}-\eta_c x_l) {\cal I} ,\label{WorkSCGen} 
\end{eqnarray}
and the thermodynamic efficiency~(\ref{MEqaaaam}) reads
\begin{equation}
\eta =\frac{\mu_l-\mu_r}{\varepsilon-\mu_r}=1-(1-\eta_c)\frac{x_l}{x_r} .\label{EfficiencySCGen}
\end{equation}
The properties of the system are contained in the dependence of the flux
$ {\cal I}$ on the variables $x_l$ and $x_r$, $ {\cal I}={\cal I}(x_r,x_l)$.  Its explicit
expression is obtained by inserting the steady state solution of the master equation~(\ref{MEq})
into the expression~(\ref{MEqaaaaf}) for the mass flux.

To identify the regime of maximum power, we proceed in two steps. The extremum of power with
respect to ${\cal F}$ is determined by the condition
\begin{eqnarray} 
T_r^{-1} \frac{\partial \dot{{\cal W}}}{\partial {\cal F}} = {\cal I}+ ({\cal F} - x_l \eta_c)
\partial_{{\cal F}} {\cal I} = 0 .
\label{maximumCond1}
\end{eqnarray}
Since we are interested in the behavior around equilibrium, including the first nonlinear correction term to
the linear regime, we expand the current ${\cal I}(x_r,x_l)= {\cal I}(x_l-{\cal F},x_l)$  to quadratic
order in  ${\cal F}$, $ {\cal I} = L {\cal F} + M {\cal F}^2 + {\cal O}({\cal F}^3)$.
The Onsager coefficients are given by $L=-{\cal I}'_1(x_l,x_l)$ and $M={\cal I}''_{11}(x_l,x_l)/2$.
The primes denote the number of derivatives and the sub-indices indicate whether these derivatives are
taken with respect to the first or second variable.
Furthermore, since ${\cal F}$ has to become zero when $\eta_c$ goes to zero, we can write, again
to quadratic order, that ${\cal F}= b_1 \eta_c +c_1 \eta_c^2 + {\cal O}(\eta_c^3) \label{y1expansion}
$.
Insertion in the extremum condition~(\ref{maximumCond1}) allows us to identify the coefficients
$b_1= {x_l}/{2}$ and $c_1={M x_l^2}/{(8L)}$.
The resulting expression  for the efficiency (\ref{EfficiencySCGen}) reads
\begin{eqnarray} 
\eta = \frac{\eta_c}{2} + (\frac{1}{4}-\frac{M x_l}{8L}) \eta_c^2 +  {\cal O}(\eta_c^3) .
\label{effexpan1}
\end{eqnarray}

Next, we maximize power with respect to $x_l$,
\begin{eqnarray} 
T_r^{-1} \frac{\partial \dot{{\cal W}}}{\partial x_l} = -\eta_c {\cal I} + ({\cal F} -
x_l \eta_c) \partial_{x_l} {\cal I} = 0 .
\label{maximumCond2}
\end{eqnarray}
It suffices to find the result to lowest order in $\eta_c$. Inserting the expansions
${\cal I}= L {\cal F}  +  {\cal O}(\eta_c^2)$ and  ${\cal F}= \eta_c x_l /2 +  {\cal O}(\eta_c^2)$,
one finds that $x_l = -{2L}/{\partial_{x_l}L}$.
Combined with Eq.~(\ref{effexpan1}), we finally arrive at the following result for the efficiency at maximum
power, valid up to quadratic order in $\eta_c$:
\begin{eqnarray} 
\label{AnalyExpEta}
\eta=\frac{\eta_c}{2}+ (1+\frac{M}{\partial_{x_l}L})\frac{\eta_c^2}{4}+ {\cal O}(\eta_c^3) .
\end{eqnarray}
We conclude that, while we recover the universal value of the coefficient $1/2$ in the linear term,
the coefficient of the quadratic term is in general  model dependent. However,
as we now proceed to show, the appearance of the coefficient $1/8$  in several previously studied
models \cite{Seifert07b,EspositoLindWdB,Tu}  derives from the fact that these models possess
a left-right symmetry. More precisely, such a symmetry implies that the switching of temperatures
$\beta_{\nu}$ and chemical potentials $\mu_{\nu}$ leads to an inversion of the flux,
\begin{equation}
{\cal I} (x_r,x_l)=-{\cal I} (x_l,x_r)\label{fluxsymmetry}.
\end{equation}
By deriving  both sides with respect to $x_r$ and $x_l$ and then setting $x_r=x_l$, one finds
that ${\cal I}''_{12}(x_l,x_l)=0$. Together with $\partial_{x_l}L=-{\cal I}''_{11}(x_l,x_l)-{\cal I}''_{12}(x_l,x_l)$,
we conclude that the condition $2M=-\partial_{x_l}L$ is verified and universality of the
coefficient $1/8$ is established under the symmetry specified in Eq.~(\ref{fluxsymmetry}).

\begin{figure}[t]
\centering
\rotatebox{0}{\scalebox{0.5}{\includegraphics{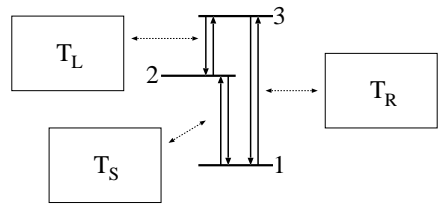}}}
\caption{Illustration of the maser model}
\label{plotmaser}
\end{figure}

To illustrate these findings, we turn to the analysis of the maser model introduced in~\cite{Scovil59},
see Fig.~\ref{plotmaser}. The system possesses three energy levels $ \epsilon_i$, $i=1,2,3$. It exchanges
photons with three equilibrium black bodies $R$, $L$ and $S$ (temperatures $T_r$, $T_l$ and $T_s$) with
corresponding specific frequencies $h \nu_r=\epsilon_3-\epsilon_1$, $h \nu_l=\epsilon_3-\epsilon_2$
and $h \nu_s=\epsilon_2-\epsilon_1$. The reservoirs  $R$, $L$ and $S$ control the transitions
$1-3$, $2-3$ and $1-2$, respectively. The stochastic dynamics of these transitions are described by the
master equation~(\ref{MEq}), with rates corresponding to the processes of absorption, spontaneous
emission, and stimulated emission of the photons.  They are given by
$W_{31}^{(r)} = \Gamma_{r} n(x_{r})$  (absorption of a photon from $R$) and
$W_{13}^{(r)} = \Gamma_{r}[1+n(x_{r})]$ (spontaneous and stimulated emission of a photon into $R$),
and identical expressions for transitions  $2-3$ and  $1-2$, with the indices $r$ replaced by $l$ and $s$ respectively.
Here, we introduced  the Bose-Einstein distribution $n(x)=[\exp(x)-1]^{-1}$, with the scaled
energies $x_r = h \nu_r/T_r$, 
$x_l = h \nu_l/T_l$ and $x_s = h \nu_s/ T_s $, and the reduced Einstein coefficients $\Gamma_{\nu}$.
 
To transform the system into a thermal engine, we consider the high temperature limit
$T_s \to \infty$ ($x_s \to 0$). The reservoir $S$  effectively becomes a repository of work, since heat
stored in a reservoir at infinite temperature can be recuperated at 100$\%$ (the corresponding Carnot
efficiency being equal to 1). We next note the cyclic nature of the transitions:
Starting from state $1$, in order to deposit the amount $-{w}=h\nu_s$ as work into the $S$-reservoir
(transition $2 \rightarrow 1$), the system first needs to absorb a photon ${q}^{(r)}=h\nu_r$ from
the hot reservoir (transition $1 \rightarrow 3$) and next deposit $-{q}^{(l)}=h\nu_l$ into the cold reservoir
(transition $3 \rightarrow 2$). This (with the reverse process) is the only available cycle.
The corresponding efficiency of the cycle reads:
\begin{eqnarray} 
\eta =\frac{-w}{{q}^{(r)}}= \frac{\epsilon_2-\epsilon_1}{\epsilon_3-\epsilon_1}
= 1-(1-\eta_c) \frac{x_l}{x_r} \label{EfficMaser}.
\end{eqnarray}
\begin{figure}[h]
\centering
\begin{tabular}{c@{\hspace{0.5cm}}c}
\hspace{0.35cm}
\rotatebox{0}{\scalebox{0.45}{\includegraphics{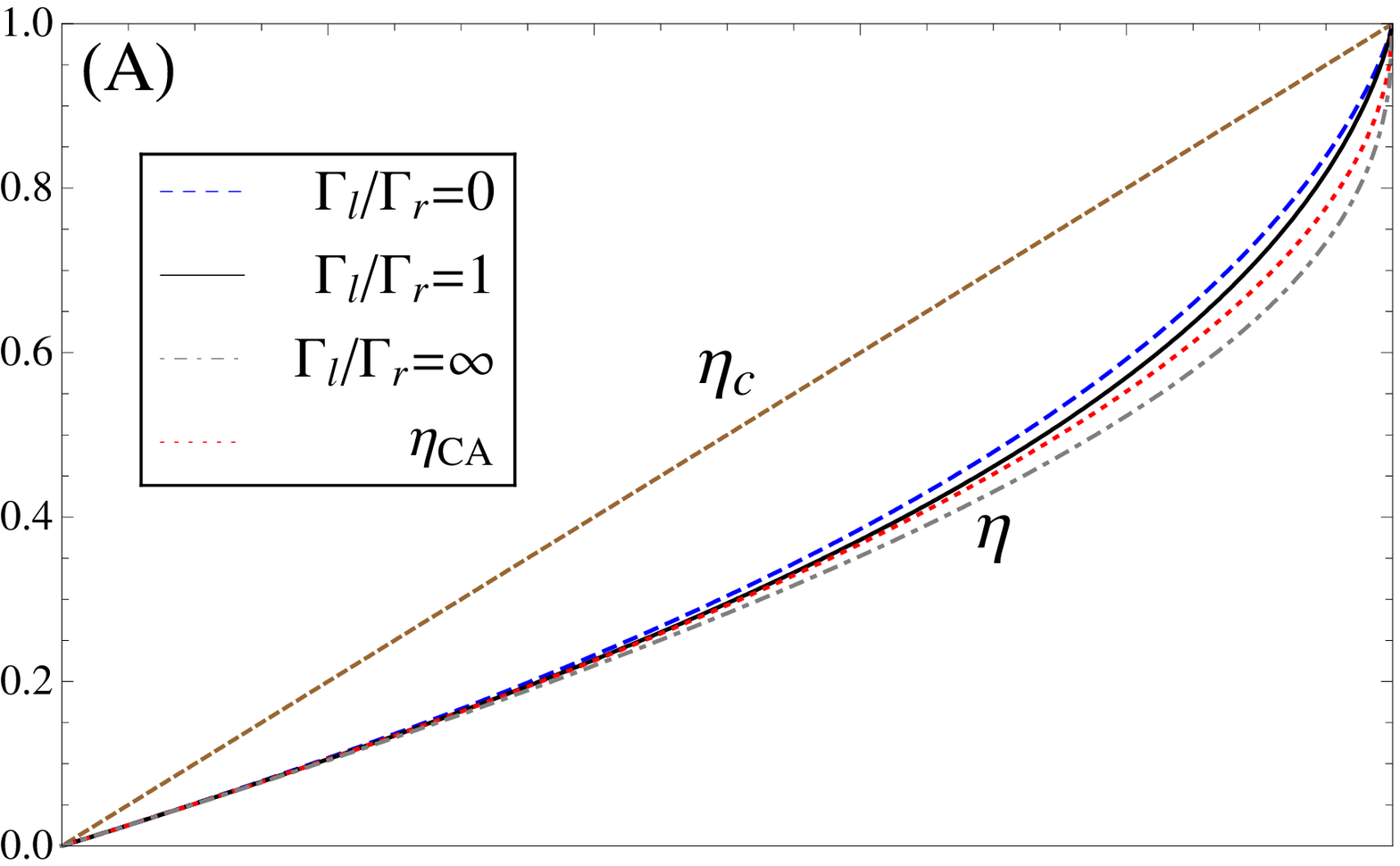}}} \vspace{-0.26cm}\\
\rotatebox{0}{\scalebox{0.45}{\includegraphics{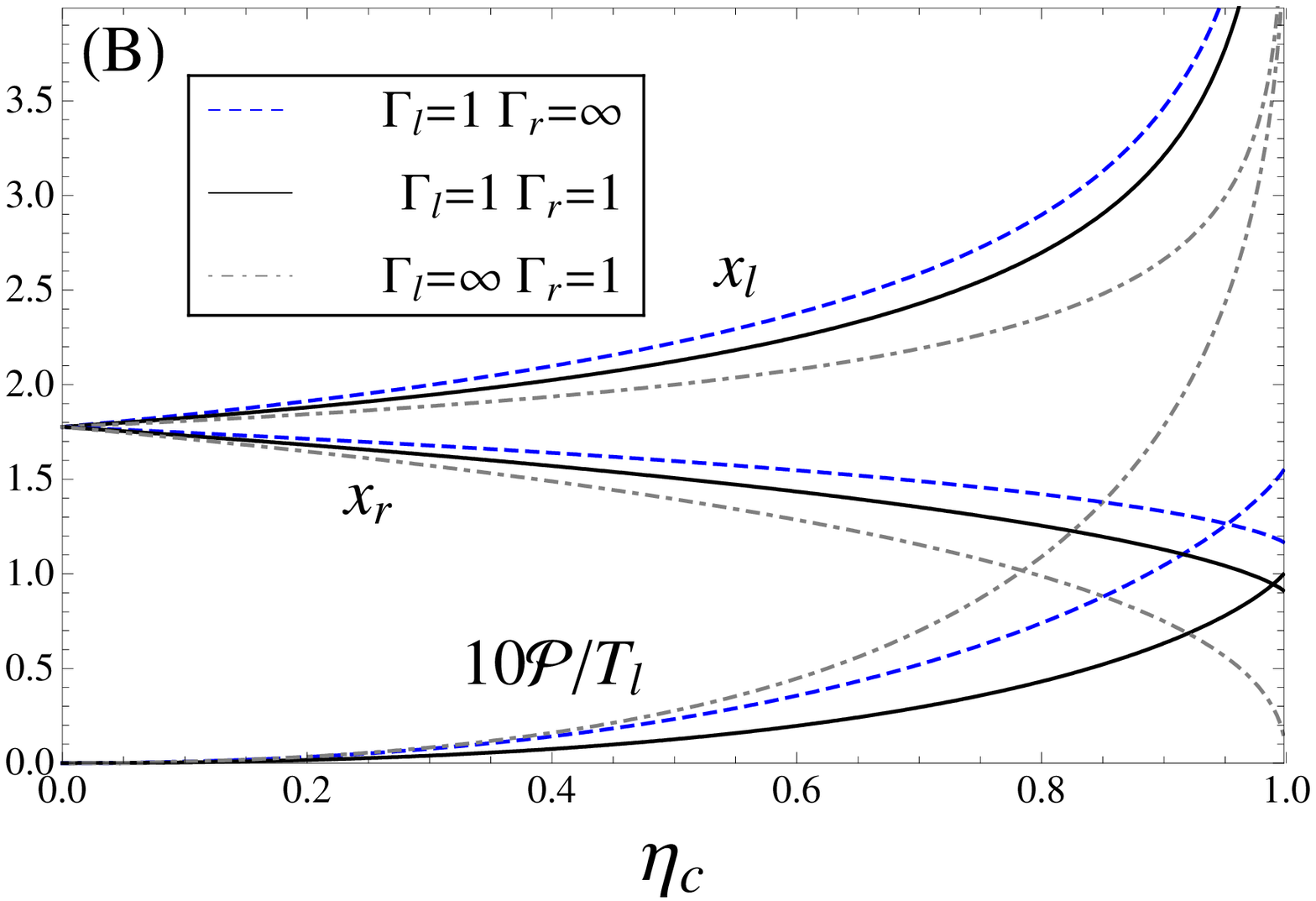}}}
\end{tabular}
\caption{(Color online) (A) Efficiency at maximum power compared with Carnot efficiency
(straight dashed line) and Curzon-Ahlborn efficiency (dotted line). (B) Scaled energies $x_l$
and $x_r$ and maximum power $\cal P$ (note that $x_l$ and $x_r$ only depend on the ratio 
$\Gamma_l/\Gamma_r$ but $\cal P$ not).}
\label{plotmaser2}
\end{figure}
At steady state, the system will, on average, run through ${\cal I}$  such cycles per unit time,
with corresponding heat flows $\dot{{\cal Q}}^{(r)}=q_r {\cal I}$ and $\dot{{\cal Q}}^{(l)}=q_l {\cal I}$. The power of the device is given by
\begin{eqnarray} 
{{\cal P}}=-w{\cal I}
= \dot{{\cal Q}}^{(r)}+\dot{{\cal Q}}^{(l)} 
= \big( T_r x_r- T_l x_l \big) {\cal I}. \label{DefWorkExtrMaser}
\end{eqnarray}
We thus recover the previously derived results of the strong coupling regime, 
cf. Eqs.~(\ref{WorkSCGen}) and (\ref{EfficiencySCGen}). To complete the analysis, we need to
evaluate the steady-state current ${\cal I}$. This is a matter of algebra, involving  the 
steady state solution of the master equation using the transition rates given above. One finds:
\begin{eqnarray} 
{\cal I} = \frac{(\e^{x_l}-\e^{x_r}) \Gamma_l \Gamma_r}
{(1+2\e^{x_l})(\e^{x_r}-1) \Gamma_l+(1+2\e^{x_r})(\e^{x_l}-1) \Gamma_r} .
\label{CurrHighTMaser}
\end{eqnarray}  
Concerning efficiency of the device at maximum power, we can now invoke the general conclusions
mentioned earlier. The symmetry  criterion~(\ref{fluxsymmetry}) for the current is only
satisfied when $\Gamma_l= \Gamma_r$. Under this condition, the efficiency at maximum power displays
the universal coefficient $1/8$ for the quadratic term, in addition to the universal
linear coefficient $1/2$. This observation is confirmed by an explicit calculation for the
model under consideration. We find:
\begin{eqnarray} 
\eta = \frac{\eta_c}{2} + \bigg(1-\frac{3(\Gamma_l-\Gamma_r)}{(\Gamma_l+\Gamma_r)(3 \cosh \alpha + \sinh \alpha)}\bigg) 
\frac{\eta_c^2}{8} +  {\cal O}(\eta_c^3),   \nonumber 
\end{eqnarray}
where $\alpha=1.77676$, solution of a transcendental equation 
$2+\alpha+2 \e^{\alpha}+2(\alpha-2)\e^{2 \alpha}=0$, 
is also the asymptotic value of $x_l$ and $x_r$ when $\eta_c \to 0$.  
To complete the picture, we have reproduced, in Fig.~\ref{plotmaser2}(A),
the efficiency at maximum power as a function of $\eta$, with $\eta\in[0,1]$, for the cases
$\Gamma_l/ \Gamma_r=0,1$ and $\infty$~\cite{comment}.  All three curves are remarkably close to the
Curzon-Ahlborn efficiency, even though the efficiency is slightly larger in the first two cases
and slightly less in the last case.   In view of their technological interest,  we also include in
Fig.~\ref{plotmaser2}(B) the corresponding maximum power and the operational conditions
of  the scaled energies $x_l$ and $x_r$.
\section*{Acknowledgments}

M. E. is supported by the FNRS Belgium (charg\'e de recherches) and 
by the government of Luxembourg (Bourse de formation recherches). 
This research is supported in part by the
National Science Foundation under grant PHY-0354937. 
We thank Dr. B. Cleuren for stimulating discussions.




\begin{thebibliography} {99}

\bibitem{curzon}
F. Curzon and B. Ahlborn, Am. J. Phys. {\bf 43}, 22 (1975).

\bibitem{VandenBroeck05} 
C. Van den Broeck, Phys. Rev. Lett. {\bf 95}, 190602, (2005).

\bibitem{kedem}
O. Kedem and S. R. Caplan, Trans. Faraday Soc. {\bf 61}, 1897 (1965).

\bibitem{VandenBroeck6} 
C. Van den Broeck,  Adv. Chem. Phys. {\bf 135}, 189 (2007).

\bibitem{allamajan}
A. E. Allahverdyan ,  R. S. Johal and G. Mahler,
Phys. Rev. E {\bf 77},  041118 (2008).

\bibitem{japanese}
Y. Izumida and K. Okuda, EPL. {\bf 83}, 60003 (2008).

\bibitem{Seifert07b}
T. Schmiedl and U. Seifert
EPL. {\bf 81}, 20003 (2008).

\bibitem{EspositoLindWdB}
M. Esposito, K. Lindenberg and C. Van den Broeck, 
arXiv:0808.0216v1.

\bibitem{Tu}
A. Gomez-Marin and  J. M. Sancho, Phys. Rev. E {\bf 74}
 062102 (2006); Z. C. Tu, J. Phys. A {\bf 41},  312003 (2008).

\bibitem{Prost99}
A. Parmeggiani, F. Julicher, A. Ajdari and J Prost,
Phys. Rev. E {\bf 60}, 2127 (1999).

\bibitem{Linke}
T.E. Humphrey, R. Newbury, R. P. Taylor and H. Linke
Phys. Rev. Lett. {\bf 89}, 116801 (2002);
T. E. Humphrey and H. Linke, 
Phys. Rev. Lett. {\bf 94}, 096601 (2005);
M. F. O'Dwyer, T. E. Humphrey and H. Linke,
Nanotechnology {\bf 17}, S338-S343 (2006). 

\bibitem{Cleuren}
B. Cleuren, C. Van den Broeck and R. Kawai, Phys. Rev. E {\bf 74}, 021117 (2006).

\bibitem{Scovil59}
H. E. D. Scovil and E. O. Schulz DuBois, 
Phys. Rev. Lett. {\bf 2}, 262 (1959).

\bibitem{Schnakenberg}
J. Schnakenberg, Rev. Mod. Phys. {\bf 48}, 571 (1976); 
Luo Jiu-Li, C. Van den Broeck, and G. Nicolis, Z. Phys. B {\bf 56}, 165 (1984). 

\bibitem{Esposito07}
M. Esposito, U. Harbola and S. Mukamel, 
Phys. Rev. E {\bf 76}, 031132 (2007). 

\bibitem{GrootMazur}
S. R. de Groot and P. Mazur, {\it Non-equilibrium thermodynamics} 
(Dover, New York, 1984). 

\bibitem{comment}
These results are obtained numerically, as they require the solution of transcendental equations.































































\end{thebibliography}
\end{document}